\documentclass[secnumarabic,elsarticle,reprint,groupedaddress]{revtex4-1}
\usepackage{verbatim}
\usepackage[english]{babel}
\usepackage{graphicx}
\usepackage{color}
\usepackage{amsmath}

\begin{document}

\title{Magnetic freezing transition in a CoO/Permalloy bilayer revealed by transverse ac susceptibility}
\author{Sergei Urazhdin and Weijie Li}
\affiliation{Department of Physics, Emory University, Atlanta, GA 30322}
\author{Lydia Novozhilova}
\affiliation{Department of Mathematics, Western Connecticut State University, Danbury, CT 06810}

\begin{abstract}

We utilize variable-temperature, variable-frequency magneto-optical transverse magnetic susceptibility technique to study the static and dynamical magnetic properties of a thin-film CoO/Permalloy bilayer. Our measurements demonstrate that in the studied system, the directional asymmetry of the hysteresis loop is associated mainly with the difference in the reversal mechanisms between the two reversed states of magnetization stabilized by the exchange-induced uniaxial anisotropy. The latter is found to be much larger than the exchange-induced unidirectional anisotropy of the ferromagnet. We also observe an abrupt variation of the frequency-dependent imaginary part of ac susceptibility near the exchange bias blocking temperature, consistent with the magnetic freezing transition inferred from the previous time-domain studies of magnetic aging in similar systems. The developed measurement approach enables precise characterization of the dynamical and static characteristics of thin-film magnetic heterostructures that can find applications in reconfigurable magnonic and neuromorphic circuits.

\end{abstract}
\maketitle

\section{Introduction}

Intense ongoing research in magnetism is motivated by a variety of phenomena resulting from the strong coupling of spin to structural, optical, and electronic properties of matter, with promising applications in sensing and information technologies~\cite{1974}. Historically, the main focus has been on the ferromagnetically (F) and ferrimagnetically ordered materials, thanks to their robust magnetic ordering that can be efficiently manipulated by the magnetic fields or spin currents, and detected by a variety of electronic and optical techniques~\cite{OHandley}. Other types of magnetic ordering have recently attracted significant attention, due to the possible advantages they can provide in downscaling and increased speed of magnetic nanodevices. For instance, the burgeoning field of antiferromagnetic (AF) spintronics is motivated by the considerably faster, compared to ferromagnets, timescales of magnetization dynamics, and their negligible susceptibility to perturbing magnetic fields~\cite{Jungwirth2016}. Frustrated magnetic systems forming spin liquids have also attracted significant attention, thanks to the rich physics associated with exotic states such as fractionalized excitations that can be hosted by such systems~\cite{Balents2010}. 

Magnetic heterostructures composed of materials with different magnetic properties provide an efficient approach to engineering the magnetic and electronic properties of thin films. Some of the classic examples include the exchange bias (EB) effect observed in F/AF bilayers~\cite{Meiklejohn56}, and giant magnetoresistance in magnetic spin valve heterostructures~\cite{Baibich1988,Binasch1989}. One of the general questions arising in this context concerns the nature of magnetism in heterostructures of thin films with different magnetic properties. In case of F/AF bilayers, the magnetic ordering of the two materials is incompatible: with a rare exception of atomically flat uncompensated surfaces of AF, the magnetic energies of both F and AF generally cannot be simultaneously minimized in their ordered state, due to the frustrated exchange interaction at their interface~\cite{Malozemoff1987}. 
The effect of the latter on the magnetization can be approximately described as a random effective magnetic field ${\mathbf H_R}$. Because of the interfacial origin of this effective field, it is generally expected to increase with decreasing film thickness $d$. For thick films, modest $H_R$ can result in the formation of multidomain states in AF and/or F~\cite{PhysRevB.66.014430}. In case of polycrystalline AF, the weakened exchange interaction at the grain boundaries may provide a natural upper limit for the size of AF domains~\cite{FulcomerCharap72,PhysRevB.59.3722}. Nevertheless, for large $H_R$ (at sufficiently small $d$), the characteristic size $l$ of the domains is expected to decrease with decreasing $d$.  For sufficiently small $d$, $l$ can become smaller than the domain wall width $\delta$ determined by the magnetic anisotropy of the material. In this limit, the magnetization state can no longer be described in terms of distinct domains separated by domain walls, and a new "Heisenberg domain state" (HDS) - magnetization state locally twisted throughout the magnetic film, to accommodate the large local random fields - is expected to emerge~\cite{Malozemoff1988}. The existence of HDS is still experimentally unverified, and its dynamic and thermodynamic properties remain unknown.

Recent measurements of the time-dependence of the magnetization state in thin-film F/AF bilayers revealed slow power-law magnetic aging at low temperatures, which appears to be universal for such systems~\cite{Urazhdin2015,Ma2016,PhysRevB.97.054402}. The dependence of aging on the magnetic history was inconsistent with the Arrhenius-type activation, instead indicating cooperative behaviors akin to the avalanche dynamics in glassy systems~\cite{PhysRevLett.86.5211}. 
Since aging was observed only for thin films, this glassy dynamics was tentatively attributed to the emergence of HDS. The relationship between HDS and glasses was further elucidated by the temperature dependence of aging in thin films of AF=NiO~\cite{PhysRevB.97.054402}. Aging was slow and independent of temperature $T$ below the exchange bias blocking temperature $T_B$, but abruptly accelerated above $T_B$. This observation, together with the analysis of the qualitative change at $T_B$ of the dependence of aging characteristics on the magnetic history, was interpreted in terms of the glass transition in the frustrated NiO film, from a fragile magnetic solid at $T<T_B$ to a correlated magnetic liquid at $T>T_B$. We note that the microscopic spin configuration of HDS is distinct from the traditional spin glasses formed by isolated randomly interacting impurity spins embedded in a non-magnetic matrix~\cite{RevModPhys.58.801}, since local magnetic correlations in this state remain similar to those in the parent magnetically-ordered state. Nevertheless, HDS was shown to exhibit the salient characteristics of glasses, so it may be appropriate to describe this state as a putative magnetic glass.
  
The results summarized above suggested the possibility to engineer frustrated magnetism in thin films, which can become useful for new applications in information technology.  However, time-domain aging measurements did not provide sufficient information for the quantitative understanding of these systems. Here, we show that a significant insight into the nature of the magnetization states in thin F/AF films can be provided by the transverse ac susceptibility technique~\cite{PhysRevB.62.8931,Rezende2003}, which enables phase-sensitive characterization of the dynamical magnetization response. We utilize this technique to show that at temperatures above the Neel temperature $T_N$ of AF=CoO, the susceptibility is real and independent of frequency, and is identical to that of a standalone ferromagnet. The susceptibility is also real at temperatures far below $T_B$, but is significantly modified by the effects of exchange interaction between F and the frozen-magnetization state of AF.

Our central result is the emergence, at temperatures close to $T_B$, of a large and strongly temperature- and frequency-dependent imaginary part of susceptibility associated with the viscous losses in AF. Analysis shows that the magnetic viscosity varies by four orders of magnitude over the temperature range of $20$~K near $T_B$, consistent with the magnetic freezing transition expected from the prior time-domain magnetic aging measurements. In addition, the ability to precisely characterize the magnetic anisotropy allowed us to establish that in the studied material system, the asymmetry of the hysteresis loop is mainly caused by the difference in the reversal mechanisms between the opposite magnetization states stabilized by the large uniaxial anisotropy induced in F by the exchange interaction with AF. The observed difference between the reversal mechanisms is likely associated with the microscopic directionality of the magnetization state formed in AF during the magnetic freezing. Surprisingly, the unidirectional anisotropy of F, which manifests in a difference between the  susceptibility in the  magnetization configurations aligned with and opposite to the cooling field, is very small in the studied system. These results shed new light on the exchange interactions at F/AF interfaces, their effects on the static and dynamical characteristics of magnetization states, and indicate a possible route for engineering these effects.

\section{Materials and Methods}\label{experiment}

In this Section, we outline the experimental technique, describe the setup utilized in our measurements, and provide details on the sample preparation.

\subsection{Transverse magnetic ac susceptibility}

Measurements of ac susceptibility have been widely utilized for the characterization of phases of matter. For instance, the low-frequency divergence of susceptibility provides a robust indication for second-order phase transitions, while its dependencies on temperature, frequency and other experimental parameters have been utilized to establish the universal laws governing these transitions~\cite{PhysRevB.4.3174}. The frequency-dependent peak observed in ac susceptibility at the glass transition temperature has been extensively studied as one of the key signatures of the underlying phenomena~\cite{RevModPhys.58.801}. It has been also suggested that glass transition may be associated with a weak logarithmic low-frequency divergence of susceptibility, which may become important for the identification and classification of frustrated states of matter~\cite{PhysRevLett.74.1230}. 

Magnetic ac susceptibility generally depends on the direction of the driving ac field~\cite{Otake1980}. In the studies of F/AF bilayers such as those described below, the F layer becomes saturated at modest magnitudes of dc field $\mathbf{H_{dc}}$, while direct coupling of the Neel order of AF to $\mathbf{H_{dc}}$ is negligible. In this configuration, a small ac field $\mathbf{h}_{ac}(t)=\sqrt{2}\mathbf{H}_{ac}\cos\omega t$ applied along $\mathbf{H_{dc}}$ does not significantly perturb the magnetic system. Here, $H_{ac}$ is the rms magnitude of ac field. Throughout this paper, we assume that the frequency $f=\omega/2\pi$ is significantly smaller than the ferromagnetic resonance frequency of F. For $\mathbf{h_{ac}}$ transverse to $\mathbf{H_{dc}}$, the magnetization $\mathbf{M}$ of F is expected to oscillate, following the direction of the total field $\mathbf{H_{dc}+h_{ac}(t)}$. The transverse ac magnetic susceptibility $\chi_T$ can be defined as the ratio of the amplitude of the ac magnetization component to the amplitude of the ac field~\cite{PhysRevB.62.8931,Rezende2003,PhysRevB.76.054409}. However, in practice, a lock-in technique is utilized to measure the ac magnetization, providing the (generally complex) harmonics $M_T^{(n)}$ of ac magnetization $m_T(t)=\sqrt{2}Re(\sum_{n} M_T^{(n)}\exp(-i\omega n t))$ selected by the lockin frequency. Below, we colloquially refer to the experimentally measured normalized first harmonic $M_T^{(1)}/H_{ac}\equiv M_T/H_{ac}$ as the transverse susceptibility. We note that this quantity does not directly reflect the response of AF to perturbations, which is the main subject of the present work, since it remains finite even in the absence of AF. Therefore, to avoid confusion about the response studied in our work, in this paper we focus on the measurements and analysis of the first harmonic $M_T$ of ac magnetization, normalized by $M$.

\begin{figure}
	\includegraphics[width=1\columnwidth]{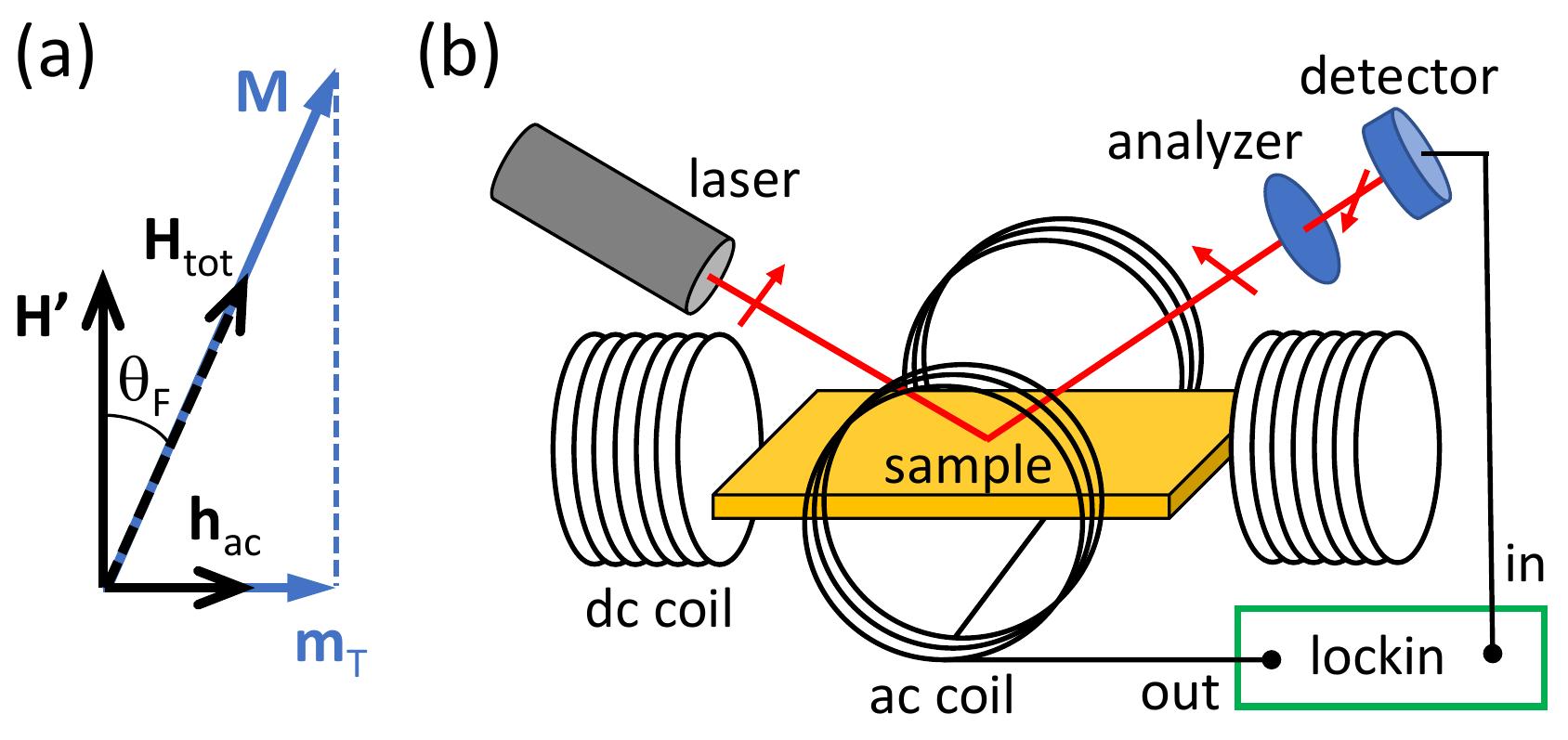} 
	\caption{\label{fig:1} (Color online). (a) Schematic of the field and the magnetization configuration in the transverse ac susceptibility measurement. Time-dependent ac field $\mathbf{h_{ac}}$ is orthogonal to the total effective dc field $\mathbf{H'}$. The oscillating component $\mathbf{m_T}$ of magnetization $\mathbf{M}$, which is aligned with the oscillating total field, is measured. (b) Schematic of the transverse ac susceptibility measurement utilizing magneto-optic Kerr effect. The sample is placed on the cold finger of the cryostat in the in-plane field of transversely oriented dc and ac coils. The laser beam is obliquely scattered from the sample surface. The oscillating Kerr rotation of polarization is measured by a combination of an analyzer and a  photodetector connected to a lockin amplifier.}
\end{figure}

The magnetization $\mathbf{M}$ of F experiences not only external magnetic fields, but also exchange interactions at the F/AF interface. By analyzing the oscillation of $\mathbf{M}$ in response to $\mathbf{h}_{ac}(t)$, one can characterize this interaction, as follows~\cite{Rezende2003,PhysRevB.68.220401}. Assume that the effect of AF on $\mathbf{M}$ can be described by an effective  time-independent unidirectional anisotropy field $\mathbf{H_{ud}}$. We can define the total effective static field $\mathbf{H'=H_{dc}+H_{ud}}$, such that $\mathbf{M}$ follows the instantaneous direction of the total effective field $\mathbf{H'+h_{ac}(t)}$, Fig.~\ref{fig:1}(a). The normalized time-dependent ac magnetization, obtained by analyzing the geometric relations illustrated in Fig.\ref{fig:1}(a), is $\frac{m_T(t)}{M}=\frac{h_{ac}}{\sqrt{h_{ac}^2+H'^2}}$. The first harmonic of ac magnetization, normalized by $M$, is given by the Fourier component of this expression,
\begin{equation}\label{MTprecise}
\frac{M_T}{M}=\frac{\omega}{\pi}\int_{-\pi/\omega}^{\pi/\omega}\frac{\cos^2\omega tdt}{\sqrt{2\cos^2\omega t+(\frac{H'}{H_{ac}})^2}}.
\end{equation}
It can be reduced to a linear combination of complete elliptic integrals of the first and second kind, which does not have an analytic expression in terms of elementary functions. We can approximate
$\frac{m_T(t)}{M}=\frac{h_{ac}}{H'}-\frac{h_{ac}^3}{2H'^3}+O(h_{ac}^5)$,  giving the first harmonic of ac magnetization 
\begin{equation}\label{MTTaylor}
\frac{M_T}{M}=\frac{H_{ac}}{H'}-\frac{3H_{ac}^3}{4H'^3}+O(H_{ac}^5). 
\end{equation}
This expansion can be replaced with
\begin{equation}\label{MT}
\frac{M_T}{M}=\frac{H_{ac}/H'}{\sqrt{1+\frac{3H_{ac}^2}{2H'^2}}}+O(H_{ac}^5).
\end{equation}
Compared to the precise Eq.~(\ref{MTprecise}), the approximation Eq.~(\ref{MT}) gives an error of less than $2\%$ for $H_{ac}<H'$. The small-amplitude limit of Eq.(\ref{MT}), at $H_{ac}\ll H'$, illustrates the usefulness of transverse susceptibility in analyzing the exchange interaction at F/AF interface. To the lowest order in $H_{ac}/H'$, we obtain 
\begin{equation}\label{MTlin}
\frac{M_T}{M}=\frac{H_{ac}}{H_{dc}+H_{ud}}+O(H_{ac}^3).
\end{equation}
Thus, in the limit $H_{ac}\ll H'$, the inverse of $M_T$ linearly depends on $H_{dc}$, and the intercept of this dependence with the field axis is $-H_{ud}$. This approximate relationship has been utilized in the previous measurements of transverse ac susceptibility to determine $H_{ud}$ in F/AF bilayers~\cite{PhysRevB.62.8931,PhysRevB.68.220401}. The linear approximation Eq.(\ref{MTlin}) deviates from the precise result Eq.~(\ref{MTprecise}) by less that $1\%$ at $H_{ac}<0.1H'$.

\subsection{Sample}

CoO(x)Co(1.5)Py(50)AlO$_x$(2) multilayers were deposited at room temperature on the oxidized Si substrate by magnetron sputtering, in an ultrahigh vacuum chamber with the base pressure of $7\times 10^{-9}$~Torr. All thicknesses are given in nanometers.  The deposition rates varied from $0.2$~A/s for AlO$_x$ to $1.8$~A/s for Py, and were calibrated using a quartz crystal microbalance.  The deposition was performed in an in-plane magnetic field $H_{dep}\approx 100$~Oe, which is empirically known to enhance EB. While this effect is commonly attributed to the magnetic ordering established during the deposition, this assumption is not applicable to CoO, whose Neel temperature $T_N$ is below room temperature. We will show below that $H_{eff}$ induces a small uniaxial anisotropy observed even above $T_N$, likely associated with magnetoelastically-induced strain in Py. This anisotropy is unrelated to the anisotropies associated with EB, as will be shown below.

The polycrystalline CoO layer was sputtered in Ar/O$_2$ mixture, using a pure Co target and the deposition parameters optimized in our previous studies of CoO-based heterostructures~\cite{Ma2016,PhysRevB.78.052403}. The other materials were sputtered from the stoichiometric targets in ultrahigh purity Ar. The soft ferromagnet Py was chosen to minimize the effects of the intrinsic magnetocrystalline anisotropy on the magnetic susceptibility. The Co(1.5) insert between CoO and Py served as an oxygen diffusion barrier preventing the oxidation of Py, which in our experience can result in a gradual deterioration of the magnetic characteristics related to EB. Optically transparent AlO$_x$ capping layer was utilized to protect the surface of Py from oxidation. 

The thickness of $x=6$~nm for the CoO layer, selected for the measurements described below, was sufficiently small so that this material was expected to form the Heisenberg domain state, as inferred from the prior time-domain studies of thickness-dependent magnetization aging~\cite{Ma2016}. On the other hand, this thickness was significantly larger than the typical CoO roughness of about $0.3$~nm rms, as measured by the atomic force microscopy, to minimize the inhomogeneous broadening of the magnetic freezing transition demonstrated below.

\subsection{Measurement technique}

Figure~\ref{fig:1}(b) shows a schematic of our transverse ac susceptibility measurement, performed in a $80-325$~K optical cryostat. Planar magneto-optical Kerr effect (MOKE) was used to detect the dynamical magnetization of the sample. An external electromagnet produced an in-plane dc magnetic field $H_{dc}$ of up to $500$~Oe, while the in-plane ac field $H_{ac}$ transverse to the dc field was produced by a coil wrapped on a slotted toroidal ferrite core built into the cryostat. The sample was mounted on a silicon cold finger, with a Cernox thermometer mounted within less than $1$~mm from the sample. The slotted ferrite core of the ac coil confined the ac magnetic field, minimizing the parasitic effects of induced emf in the metallic parts of the cryostat, as well as the Faraday effect in the optical cryostat window. At room temperature $T=295$~K, this setup allowed us to achieve ac fields of more than $100$~Oe at frequencies $f$ of a few Hz, and over $25$~Oe at the maximum measurement frequency $f=7$~kHz. The maximum achievable ac field was somewhat reduced at lower $T$, due to the decreased magnetic permeability of ferrite. The amplitude and the phase of the field produced by the coil were calibrated to the precision of $\pm2\%$ over the temperature and frequency ranges of our measurements, by utilizing a $50$~nm-thick Py film whose response followed Eq.~(\ref{MT}) with $H'=H_{dc}$. 

Magneto-optical measurement of the dynamical magnetization state utilized a polarized HeNe laser spot at a $60^\circ$ angle of incidence. The polarization rotation resulting from the planar MOKE effect was periodically modulated by the oscillating magnetization. This rotation was converted into the intensity oscillations by an analyzer. The oscillations of photocurrent produced by the Si photodetector were measured by the phase-sensitive lockin amplifier. We note that the longitudinal MOKE effect, determined by the projection of magnetization on the dc field, may also contribute to the polarization rotation, but by the symmetry of our measurement geometry, this effect produced only $2f$ and higher harmonics that were rejected by the lockin.

\section{Results}

\subsection{Transverse ac susceptibility at $T>T_N$}\label{chihighT}

\begin{figure}
	\includegraphics[width=1\columnwidth]{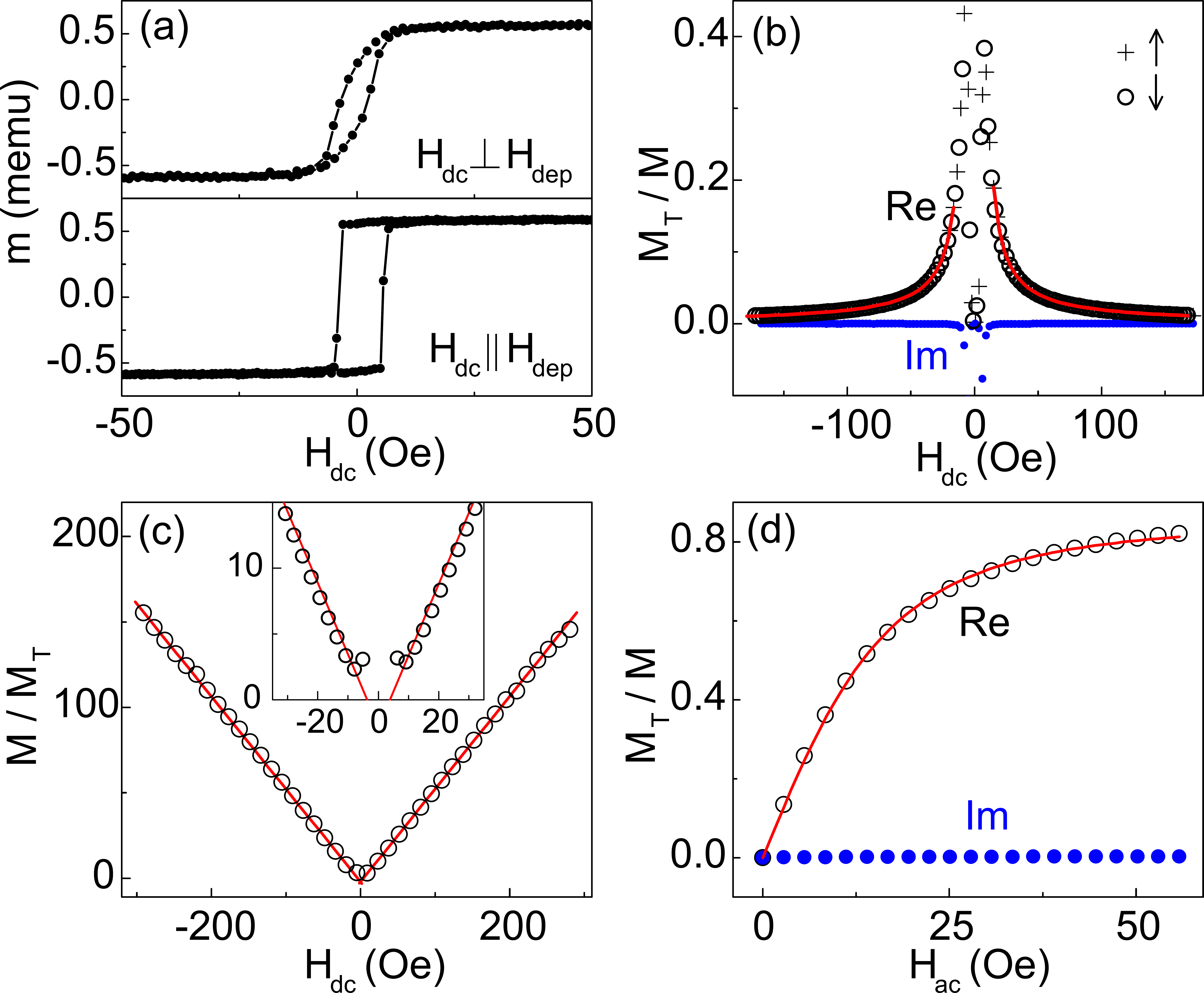} 
	\caption{\label{fig:2} (Color online). Results of measurements at $T=320$~K, above the Neel temperature of CoO. (a) Magnetic hysteresis loops with in-plane field oriented perpendicular (top) and parallel (bottom) to the deposition field. (b) Real and imaginary parts of normalized ac magnetization vs field, as labeled, measured at $H_{ac}=1.6$~Oe, $f=1$~kHz. Note that the value of the real part for the upward field scan (crosses) and for the downward scan (circles) coincide. Curves show fits of data with Eq.~(\ref{MT}), with $H'=H_{dc}+H_{ua}$ at $H_{dc}>0$, $H'=H_{dc}-H_{ua}$ at $H_{dc}<0$, and the best-fit value $H_{ua}=-5.6\pm 0.1$~Oe. (c) Symbols: inverse of Re($M_T$/M) from panel (b), lines: linear fits. Inset: zoom-in. (d) Real (open symbols) and imaginary (solid symbols) parts of normalized transverse magnetization vs ac field, at $H_{dc}=35.7$~Oe, $f=1$~kHz. Curve is a fit with Eq.~(\ref{MTua}), yielding $H_{ua}=-7.5\pm 0.2$~Oe.}
\end{figure}

At temperatures $T$ above the Neel temperature $T_N=291$~K of CoO, this layer is not expected to affect the transverse susceptibility of Py, allowing us to validate the transverse ac susceptibility technique, and establish its effectiveness in the quantitative characterization of magnetic properties. Figure~\ref{fig:2} shows the results for $T=320$~K. The magnetic hysteresis loop, obtained using a vibrating sample magnetometer (VSM), is tilted for the in-plane field $H_{dc}$ oriented perpendicular to the bilayer deposition field $H_{dep}$, but is square for $H_{dc}\parallel H_{dep}$ [Fig.~\ref{fig:2}(a)]. These behaviors indicate a small in-plane uniaxial anisotropy of magnetoelastic origin, acquired by Py due to the strain of the same origin produced by in-field deposition.

In measurements of ac susceptibility, performed with the field $\mathbf{H_{dc}}\perp \mathbf{H_{dep}}$, the normalized real part of ac magnetization $Re(M_T/M)$ rapidly increases with decreasing magnitude of $H_{dc}$, exhibits peaks at $H_{dc}=9$~Oe and $-9$~Oe for both directions of the field sweep, and reduces to zero at $H_{dc}=0$. Meanwhile, the imaginary part $Im(M_T/M)$ remains negligible at all $H_{dc}$, except for the sharp peaks at $H_{dc}=\pm 9$~Oe coinciding with the peaks of the real part. These data were acquired at $f=1$~kHz; the results were independent of $f$ over the accessible range. The inverse of $Re(M_T/M)$ is well approximated by two linear dependencies [Fig.~\ref{fig:2}(c)], intercepting the field axis at the opposite values for opposite field directions [inset in Fig.~\ref{fig:2}(c)]. These intercepts cannot be attributed to the exchange bias, since according to Eq.~(\ref{MTlin}), effective exchange-bias field $H_{ud}$ would have produced the same intercept for both field directions, rather than two opposite offsets.

We now show that the observed behaviors can be explained by the uniaxial anisotropy acquired by Py due to the in-field deposition, consistent with the hysteresis loops in Fig.~\ref{fig:2}(a). We also demonstrate how the transverse ac susceptibility technique allows quantitative characterization of this anisotropy. We start with a simplified   analysis of the magnetization state of Py in the absence of ac field and of the exchange bias from CoO. The expression for the magnetic energy density of Py then includes only the Zeeman and the uniaxial anisotropy terms~\cite{Stoner1948},
\begin{equation}\label{energy}
u=-MH_{dc}\cos(\theta)-MH_{ua}\cos^2(\theta)/2.
\end{equation}
Here, $\theta$ is the angle formed by the Py magnetization $\mathbf{M}$ relative to the dc field [see Fig.~\ref{fig:1}(a)], and $H_{ua}$ is the effective uniaxial anisotropy field, defined to be positive for $H_{dc}$ parallel to the easy axis, and negative if $H_{dc}$ is perpendicular to it. To the lowest order in $\theta$, $u\approx -MH_{dc}\theta^2/2-MH_{ua}\theta^2/2$+const. The effect of uniaxial anisotropy at small $\theta$ is thus equivalent to that of the field $H_{ua}$ for $H_{dc}>0$, and $-H_{ua}$ for $H_{dc}<0$. Indeed, Eq.~(\ref{MT}) with $H'=H_{dc}+H_{ua}$ at $H_{dc}>0$, $H'=H_{dc}-H_{ua}$ at $H_{dc}<0$, and the best-fit value $H_{ua}=-5.6\pm 0.1$~Oe provides an excellent fit to the high-field data [curves in Fig.~\ref{fig:2}(b)]. Note that the negative value of $H_{ua}$, determined from the fitting, means that the direction of easy axis is perpendicular to $H_{dc}$, consistent with the magnetometry data in Fig.~\ref{fig:2}(a).

The peaks in ac susceptibility at $H_{dc}$ approaching $-H_{ua}$ arise because the response to $\mathbf{h_{ac}}$, which to the lowest order in $\theta$ is determined by the inverse of the derivative $\frac{\partial u}{\partial\theta}(\theta=0)$, diverges at $H_{dc}=-H_{ua}$. This divergence can be observed only for the dc field perpendicular to the easy axis, since the relation $H_{dc}=-H_{ua}$ requires that $H_{ua}<0$ at $H_{dc}>0$, and vice versa. To interpret the peaks in the imaginary part of susceptibility, we note that the unavoidable local variations of anisotropy result in an inhomogeneous magnetization state at $H_{dc}$ close to $-H_{ua}$, giving rise to dynamical losses associated with irreversible domain wall motion. At $H_{dc}=0$, the magnetization is expected to rotate into the easy-axis direction, which in our ac susceptibility measurement geometry is perpendicular to the dc field and parallel to the ac field. As discussed above, the ac susceptibility is then expected to vanish, consistent with the data in Fig.~\ref{fig:2}(b). 

We can generalize our analysis of ac susceptibility at small $H_{ac}$ to include both the uniaxial and the unidirectional anisotropies,
\begin{equation}\label{Heff}
H_{ua}=(H_--H_+)/2,H_{ud}=-(H_++H_-)/2,
\end{equation}
where $H_+$, $H_-$ are the intercepts of the linear fits of $M_T/M$ vs $H$ for $H_{dc}>0$ and $H_{dc}<0$, respectively. If $H_+$ and $H_-$ were positive and negative coercive fields, respectively, then Eqs.~(\ref{Heff}) would become identical, up to a sign, to the definitions of the coercive field $H_C$ and the exchange bias field $H_E$ commonly introduced in the studies of exchange bias~\cite{NOGUES1999203,BERKOWITZ1999552}. Surprisingly, in the studied system these two sets of characteristics are not directly related to each other, as shown below.

\subsection{Nonlinear transverse ac susceptibility}

A complementary approach to analyzing magnetic anisotropy based on the magnetic susceptibility measurements is provided by the nonlinear regime of Eq.~(\ref{MT}), as follows. Consider the dependence of $M_T/M$ on $H_{ac}$. According to Eq.~(\ref{MTlin}), this dependence is linear at $H_{ac}\ll H'$, with the slope $1/H'$. This slope by itself cannot be used to independently determine $H'$, because the overall scaling of the measurement still needs to be determined from the dependence on dc field such as Fig.~\ref{fig:2}(c). However, at $H_{ac}$ comparable to $H'$, the dependence is expected to start saturating, because the maximim possible magnitude of $m_T$ is limited by $M$. To the lowest order, this nonlinearity is described by the cubic in $H_{ac}$ term in Eq.~(\ref{MTTaylor}). The ratio of the cubic to the linear terms in Eq.~(\ref{MTTaylor}) provides a quantitative measure of $H_{ac}/H'$, regardless of the overall data scaling, and independent from the results of linear-regime fitting discussed above. 

To account for the uniaxial anisotropy in the analysis of these nonlinear effects, we need to consider the full energy function
\begin{equation}\label{fullenergy}
u(\theta)/M=-H'\cos\theta-H_{ua}\cos^2\theta/2-h_{ac}\sin\theta,
\end{equation}
which includes the Zeeman contributions of the total effective dc field and the ac field, as well as the uniaxial anisotropy energy. The dependence $\theta(h_{ac})$ is determined by minimizing this function,
\begin{equation}\label{theta}
0=\frac{1}{M}\frac{du}{d\theta}=H'\sin\theta+H_{ua}\sin 2\theta/2-h_{ac}\cos\theta.
\end{equation}
This equation transforms into a quartic algebraic equation with respect to $\sin\theta$, whose exact solution does not provide a practical approach for our analysis. We now develop a perturbative approach allowing us to determine the expansion of $M_T/M$ in powers of $H_{ac}$, without solving Eq.~(\ref{theta}). 
To simplify the notations, we introduce $x=\sin\theta$, $y=\cos\theta$. Equation~(\ref{theta}) can be then re-written as
\begin{equation}\label{xy}
\begin{cases}
x^2+y^2=1,\\
H'x-h_{ac}y+H_{ua}xy=0.
\end{cases}
\end{equation}
By successively differentiating this system of equations with respect to $h_{ac}$, we can find any derivative of $x(h_{ac})$, $y(h_{ac})$ at $h_{ac}=0$, without having to solve these equations. Assuming for concreteness $H'>-H_{ua}$, we obtain 

\begin{align*}
x^{(2n)}(0) & =y^{(2n+1)}(0)=0,\\
y(0) & =1,\\
x'(0) & =H'/(H'+H_{ua}),\\
y''(0) & =-H'^2/(H'+H_{ua})^2,\\
x'''(0) & =-3H'^3(H'-H_{ua})/(H'+H_{ua})^4.
\end{align*}

The Taylor expansion for  $x=\frac{m_T(t)}{M}$ is thus
$$\frac{m_T(t)}{M}=\frac{h_{ac}}{H'+H_{ua}}-\frac{(h_{ac})^3(H'-H_{ua})}{2H'^3(H'+H_{ua})}+O(h_{ac}^5),$$
giving the Fourier harmonic
\begin{equation}\label{MTTaylorua}
\frac{M_T}{M}=\frac{H_{ac}}{H'}-\frac{3(H_{ac})^3(H'-H_{ua})}{4H'^3(H'+H_{ua})}+O(H_{ac}^5), 
\end{equation}
or equivalently
\begin{equation}\label{MTua}
\frac{M_T}{M}=\frac{\frac{H_{ac}}{H'+H_{ua}}}{\sqrt{1+\frac{3H_{ac}^2}{2}\frac{H'-H_{ua}}{(H'+H_{ua})^3}}}+O(H_{ac}^5),
\end{equation}
which reduces to Eq.~(\ref{MT}) in the limit $H_{ua}=0$. Equation~(\ref{MTua}) confirms that to the lowest order in $H_{ua}/H'$, the effect of uniaxial anisotropy on ac susceptibility is equivalent to that of an effective field $H_{ua}$. The next-order correction, arising from the dependence of the first term in the denominator on $H_{ua}$, can be interpreted as rescaling of $M_T/M$ by $\sqrt{\frac{H'+H_{ua}}{H'-H_{ua}}}$, and simultaneously a renormalization of the total effective field. In particular, the nonlinearity of the dependence becomes enhanced (suppressed) for $H_{ua}<0$ ($H_{ua}>0$), yielding a smaller (larger) value of the effective field extracted from fitting  Eq.~(\ref{MT}), and thus an overestimation of the magnitude of $H_{ua}$. This effect is illustrated by the analysis of the dependence on ac field shown in Fig.~\ref{fig:2}(d). The measurement was performed at $H_{dc}=35.7$~Oe, ensuring a uniform state of magnetization [see Fig.~\ref{fig:2}(a)], as also confirmed by the negligible imaginary part of ac susceptibility. The real part linearly increases at small $H_{ac}$, but at large $H_{ac}$ starts to saturate, in agreement with our analysis of the nonlinear effects. Fitting with the dependence Eq.~(\ref{MTua}) yields $H_{ua}=7.5\pm 0.2$~Oe. A minor discrepancy with the results for the dependence on $H_{dc}$ discussed above may be associated with the higher-order corrections neglected in our analysis, or with the spatial inhomogeneity of the magnetic properties neglected in our analysis. Meanwhile, fitting with Eq.~(\ref{MT}) gives $H'=25.4\pm0.2$~Oe, i.e.  $H_{ua}=10.3$~Oe, significantly overestimating the uniaxial anisotropy, in agreement with the analysis above.

The quantitative analysis discussed below will rely predominantly on the small-angle dependence Eqs.~(\ref{Heff}), unaffected by the subtle nonlinear effects associated with uniaxial anisotropy. However, such measurements are difficult in the regimes where the system exhibits a significant dependence of the magnetic properties on the magnetic history and time. Since both Eq.~(\ref{MT}) and the more precise (Eq.~(\ref{MTua})) give the same functional form of $M_T/M$, in such cases we will for simplicity analyze $H'$ determined from Eq.~(\ref{MT}) to determine, keeping in mind that to obtain $H_{ua}$, one needs to account for the corrections described by Eq.~(\ref{MTua}).

\subsection{Transverse ac susceptibility at $T<T_N$}

\begin{figure}
\includegraphics[width=1\columnwidth]{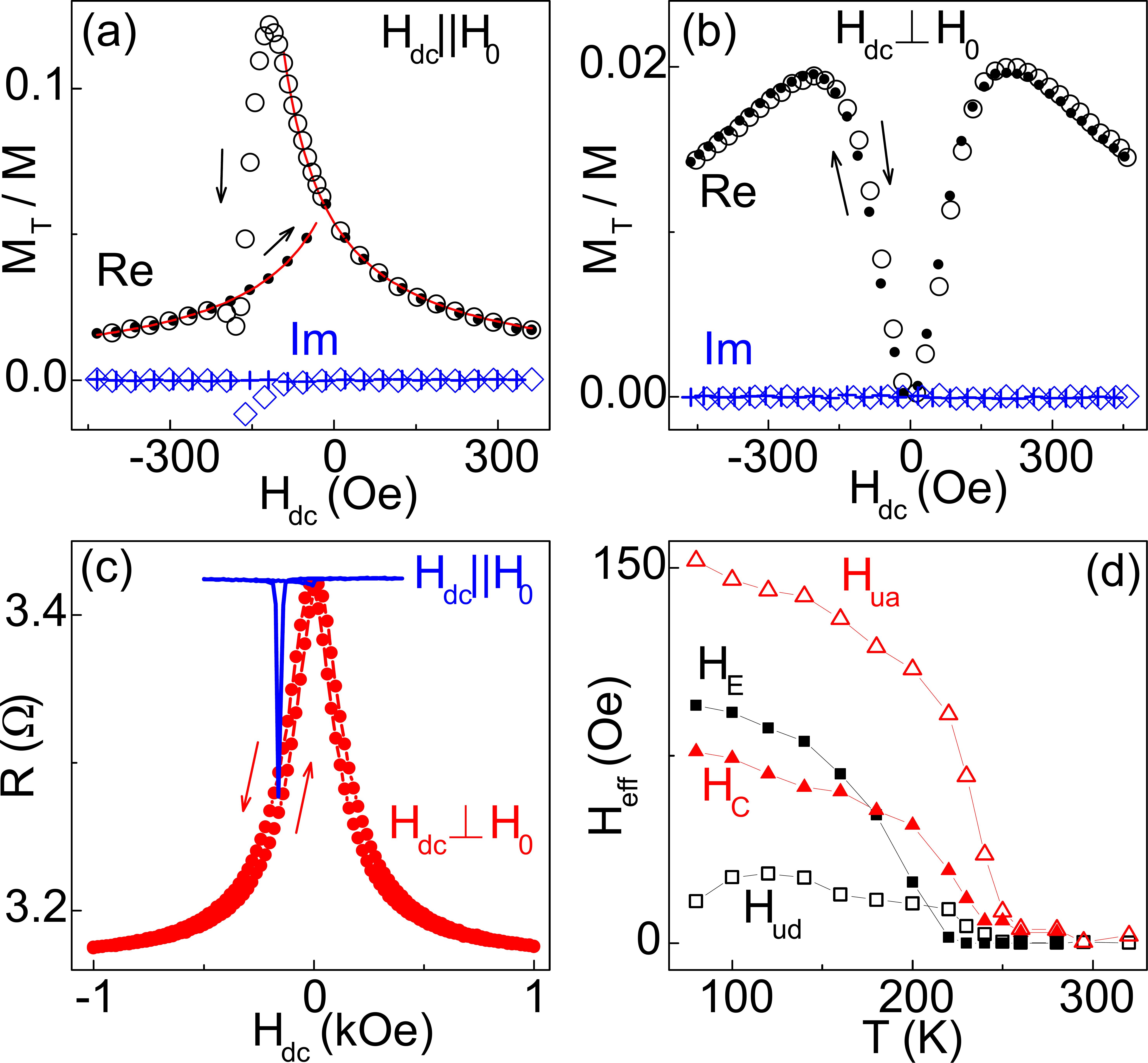} 
\caption{\label{fig:3} (Color online). (a,b) Real (circles, dots) and imaginary (crosses, diamonds) parts of $M_T/M$ vs $H_{dc}$, measured with $H_{dc}$ parallel (a) and perpendicular (b) to the cooling field $H_0$, at $T=80$~K, $H_{ac}=8$~Oe. Dots and crosses: upward field sweep, open circles and diamonds: downward sweep. Curves in (a) are the results of fitting using Eq.~(\ref{MTlin}). (c) Anisotropic magnetoresistance vs in-plane field oriented parallel and perpendicular to the cooling field $H_0$, as labeled, at $T=80$~K, with the electrical current flowing parallel to $H_0$. Measurements were performed in the van der Pauw geometry, with the four contacts attached to the corners of the chip. An ac current of $0.1$~mA rms and frequency $f=1.3$~kHz was applied to the sample, and the resulting ac voltage was detected by the standard lock-in detection technique. (d) Effective unidirectional field $H_{ud}$ (open squares), uniaxial field $H_{ua}$ (open triangles), exchange bias field $H_{E}$ (solid squares), and coercive field $H_{C}$ (solid triangles) vs temperature, determined by fitting the data such as shown in panel (a) with Eq.~(\ref{MTlin}), and then using Eq.~(\ref{Heff}).}
\end{figure}

The results for $T>T_N$ presented above confirm that the transverse ac susceptibility technique enables precise characterization of anisotropy, by analyzing its dependence on dc or ac field. Here,  we show how this approach can be utilized to gain insight into the nature of the magnetization states in F/AF bilayers below the Neel temperature of AF, $T_N=291$~K for CoO. In all of the measurements described below, the sample was cooled from $T=320$~K at a rate of about $0.5$~K/s in the magnetic field $H_0=160$~Oe. This field was oriented in-plane perpendicular to the deposition field $H_{dep}$, unless specified otherwise. 

We start with the simpler behaviors observed at sufficiently low temperatures. Figure~\ref{fig:3}(a) shows a representative dependence of $M_T/M$ on $H_{dc}$, at $T=80$~K. The overall trends are qualitatively similar to the high-temperature results shown in Fig.~\ref{fig:2}(b), with a notable exception of asymmetry with respect to the field direction, as expected due to EB. The values of the reversal fields, determined from these data, are $H_1=-14$~Oe and $H_2=-161$~Oe for the upward and downward field sweeps, respectively, yielding the effective EB field $H_E=-(H_1+H_2)/2=88$~Oe and coercivity $H_C=(H_1-H_2)/2=73$~Oe.

Surprisingly, aside from the hysteresis in the range of fields between $H_1$, $H_2$, the dependence $M_T(H_{dc})$ is almost symmetric with respect to the field reversal, implying that the effective unidirectional anisotropy field $H_{ud}$ is small [see Eq.~(\ref{Heff})]. Indeed, fitting with Eq.~(\ref{MTlin}) [curves in Fig.~\ref{fig:3}(a)] yields $H_{ud}=17\pm 0.3$~Oe, five times smaller than $H_E$. Meanwhile, $H_{ua}=153\pm 0.3$~Oe is larger than $H_{ud}$ by an order of magnitude, and is twice as large as $H_C$. Note that the imaginary component of susceptibility remains negligible at all fields, except for a small dip at the reversal, indicating that the losses associated with the irreversible domain wall motion are negligible, and the magnetization remains in a uniform state. We also note that the positive value of $H_{ua}$ means that the easy uniaxial anisotropy axis is collinear with the cooling field $H_0$ (and the measurement field $H_{dc}$), but orthogonal to the easy axis of weak uniaxial anisotropy observed at $T=320$~K [see Fig.~\ref{fig:2}]. The latter likely originates from the dopsition field-induced strain, and is unrelated to the exchange-induced anisotropy, as evident from the additional measurements discussed below.

To test our interpretation of the data in Fig.~\ref{fig:3}(a), we performed two additional sets of measurements, as illustrated in Figs.~\ref{fig:3}(b,c). First, we cooled the sample in an in-plane field $\mathbf{H_0}$ orthogonal to $\mathbf{H_{dc}}$. In this case, the behavior of ac susceptibility [Fig.~\ref{fig:3}(b)] is qualitatively different from that in Fig.~\ref{fig:3}(b). The real component is symmetric with respect to the field direction, and does not exhibit a hysteresis. At large fields, it increases with decreasing magnitude of field, peaks at $|H_{dc}|=205$~Oe, and vanishes at $H=0$. The imaginary part of susceptibility remains negligible, indicating that the magnetization state remains homogeneous. Comparing these data to Fig.~\ref{fig:2}(b) and the analysis in Section~\ref{chihighT}, we conclude that the observed behavior is consistent with the uniaxial anisotropy whose easy axis is orthogonal to $H_{dc}$, and magnitude, inferred from the positions of the peaks, is in overall agreement with our analysis of Fig.~\ref{fig:3}(a). These data also confirm that the low-temperature anisotropy is defined entirely by the cooling field, and is unrelated to the small uniaxial anisotropy of magnetoelastic origin observed above $T_N$.

To further test our interpretation of the observed behaviors, we performed additional characterization based on the anisotropic magnetoresistance (AMR) exhibited by Py, allowing us to analyze the uniformity of its field-dependent magnetization state. A similar magnetolectronic approach was utilized by us in measurements of magnetic aging in similar systems, and described in detail in Refs.~\cite{Urazhdin2015,Ma2016,PhysRevB.97.054402}.  Because of AMR, the resistance $R$ of Py exhibits a $180^\circ$-periodic dependence on the angle $\phi$ between the magnetization and the direction of current, characterized by the maximum value $R_{max}=3.425$~$\Omega$ in the studied sample at $\phi=0,180^\circ$, $T=80$~K, and the minimum value $R_{min}=3.17$~$\Omega$ at $\phi=\pm90^\circ$. When sweeping the field $H_{dc}$ collinear with the direction of the cooling field $H_0$ (and of the current in our measurement configuration), the value of $R$ remains constant and very close to $R_{max}$ at all fields, except for a sharp dip at $H_{dc}=-160$~Oe for the downward field sweep, and a smaller one at $H_{dc}=-10$~Oe for the upward field sweep, both of which can be identified with the magnetization reversal. These data demonstrate that the magnetization of Py remains close to a uniform state in the range of fields utilized in our analysis of susceptibility in panel (a). A further confirmation for the interpretation of the data in panels (a,b) in terms of a well-defined induced anisotropy is provided by the results for $\mathbf{H_{dc}}\perp \mathbf{H_0}$, also shown in Fig.\ref{fig:3}(c). The resistance is close to $R_{min}$ at large fields, gradually increases with decreasing $|H_{dc}|$, and reaches precisely $R_{max}$ at $H_{dc}=0$. Thus, at $H_{dc}=0$, the magnetization becomes uniformly orthogonal to the direction of the swept dc field, indicating a well-defined direction of the induced magnetic anisotropy,  spatially homogeneous on the lengthscale defined by the magnetic coherence length of the ferromagnet.

The lack of a direct relationship between unidirectional anisotropy and exchange bias in the studied CoO/Py bilayer is confirmed by the temperature dependence of the magnetic characteristics, determined from the hysteresis loops of $M_T/M$ acquired at temperatures between $80$~K and $320$~K, Fig.~\ref{fig:3}(d). The coercivity $H_C$, the exchange bias field $H_E$, and the effective uniaxial anisotropy field $H_{ua}$ monotonically decrease with increasing $T$, Fig.~\ref{fig:3}(b). The exchange bias vanishes above $220$~K, but the enhanced coercivity and uniaxial anisotropy persist up to $240-250$~K. Surprisingly, the unidirectional anisotropy $H_{ud}$ remains small and almost constant up to $220$~K. These results clearly demonstrate that $H_{E}$ is not correlated with the unidirectional anisotropy, instead mostly following the uniaxial anisotropy.

We now discuss the origin of the surprisingly large discrepancy between the observed large asymmetry of the magnetization reversal fields and the small unidirectional anisotropy. The underlying mechanism is elucidated by the detailed inspection of Fig.~\ref{fig:3}(a). In the downward field sweep, $M_T/M$ grows to $0.12$ at $H_{dc}=-114$~Oe, before rapidly decreasing due to the magnetization reversal. In contrast, in the upward field sweep, reversal occurs at the point of intersection between the positive- and negative-field dependencies at a small field, corresponding to a significantly smaller $M_T/M=0.06$. Thus, this reversal occurs as soon as the energy of the reversed state becomes lower, i.e. the reversal is not deterred by the energy barrier associated with the uniaxial anisotropy. One of the common mechanisms that can accomplish this is reversal via domain wall propagation, where the energy cost of the domain wall, proportional to its length, can be significantly smaller than the  energy of coherent magnetization rotation proportional to the area of the domain. Similar effects are commonly encountered in ferromagnetic films with perpendicular magnetic anisotropy (PMA), which can exhibit very small coercivity even if the PMA is large, due to the nucleation of reversed domains on defects.

The large values reached by $M_T/M$ in the downward field sweep clearly demonstrate that  reversal occurs only when the effective uniaxial anisotropy field is nearly compensated by $H_{dc}$ [see Eq.~\ref{MTTaylorua}]. In other words, the domain walls that could facilitate reversal are suppressed in this state. We conclude that in the studied system, the asymmetry between the reversal fields is dominated by the difference between the magnetization reversal mechanisms between the two opposite magnetization orientations stabilized by a large uniaxial anisotropy. The reversal asymmetry must be associated with the directionality of the stable magnetization state of AF defined by the field cooling. However, our measurements reveal that the dominant effects of this directionality are not described by  an induced unidirectional anisotropy.

Our observation of uniaxial anisotropy, with the easy axis determined by the field-cooling, is consistent with the predicted reversible spin-flop of the AF magnetic sublattices towards $\mathbf{M}$, with the AF Neel vector frozen by the field-cooling in the direction perpendicular to $\mathbf{M}$~\cite{Koon97,Schulthess98}. The asymmetry of reversal mechanisms, inferred from the ac susceptibility, is consistent with the previously observed asymmetry of the spatial characteristics of magnetization reversal in F/AF bilayers~\cite{doi:10.1063/1.1562732}.

\begin{figure}
\includegraphics[width=1\columnwidth]{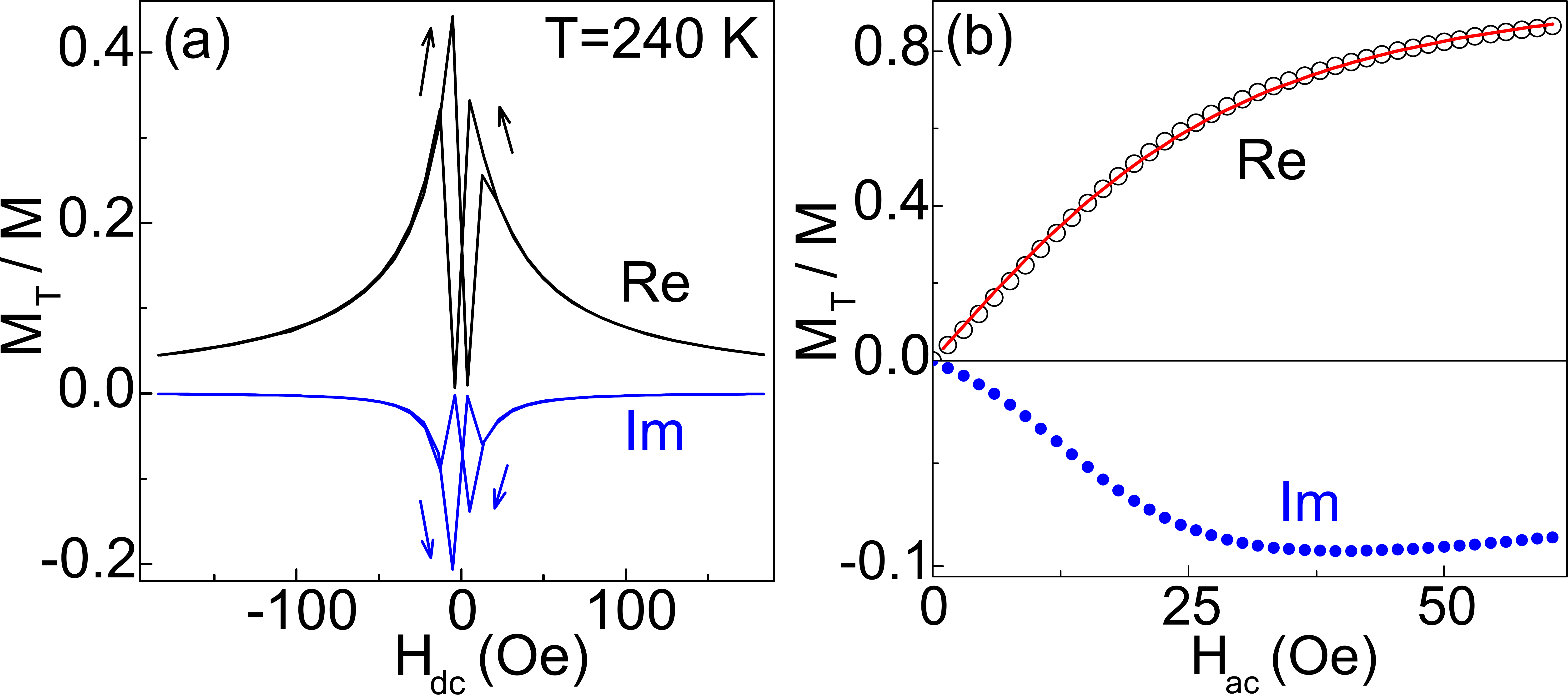} 
\caption{\label{fig:4} (Color online). Results of measurements at $T=240$~K. (a) Real and imaginary parts of $M_T/M$ vs $H_{dc}$, $H_{ac}=8$~Oe. (b) Symbols: real and imaginary parts of $M_T/M$ vs $H_{ac}$, as labeled, at $H_{dc}=35.7$~Oe. All the data were acquired at $f=1$~kHz.}
\end{figure}

We now discuss the behaviors of ac susceptibility close to the exchange bias blocking. In this temperature range, ac magnetization  exhibits a sizable imaginary part, as illustrated in Fig.~\ref{fig:4} for $T=240$~K. The variations of $Im(M_T/M)$ with $H_{dc}$ in Fig.~\ref{fig:3}(a) appear to simply mirror $Re(M_T/M)$, but the relation between these two parts is more complicated, as shown by the dependence on the ac field in Fig.~\ref{fig:4}(b). While $Re(M_T/M)$ monotonically increases with increasing $H_{ac}$, the magnitude of the imaginary part exhibits a shallow maximum at $H_{ac}=38$~Oe.

Since the low-temperature data [Fig.~\ref{fig:3}] clearly demonstrated that the magnetization of Py remains in a uniform state at fields comparable to those in Fig.~\ref{fig:4}, it is unlikely that the large imaginary component of ac magnetization originates from the irreversible domain wall motion in Py. Instead, it is likely caused by the viscous dynamics of AF magnetization associated with the previously demonstrated rotatable AF-induced anisotropy~\cite{PhysRevB.58.8605,PhysRevB.70.094420,PhysRevB.71.220410}, producing a dynamical effective exchange field exerted on $\mathbf{M}$, whose phase lags behind that of $h_{ac}$. The remainder of this paper is focused on characterizing this dynamics and analyzing its implications for the nature of the magnetization state of AF. 

\subsection{Dependence of transverse ac susceptibility on frequency}\label{chivsf}

\begin{figure}
	\includegraphics[width=1\columnwidth]{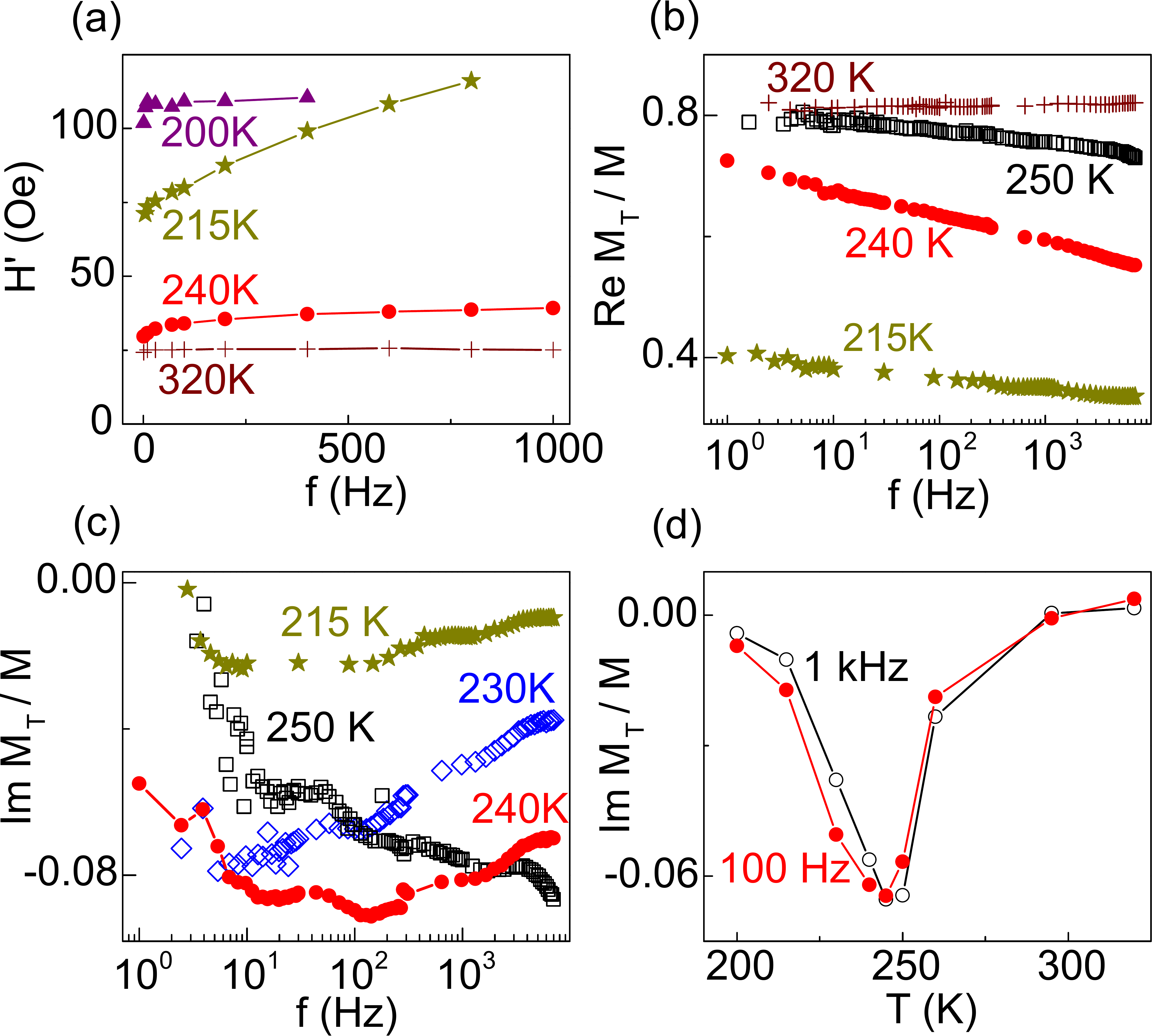} 
	\caption{\label{fig:5} (Color online). (a) Dependence of the effective field $H'$ on the frequency $f$ of ac field, at the labeled values of temperature. The values of $H'$ were determined by fitting the dependence of $Re(M_T/M)$ on $H_{ac}$ with Eq.~(\ref{MT}), at $H_{dc}=36$~Oe. (b),(c) Dependence of real (b) and imaginary (c) parts of normalized ac magnetization on $f$, at the labeled values of $T$. (d) $Im(M_T/M)$ vs $T$, at the labeled values of frequency. The data in (b)-(d) were obtained at $H_{dc}=35.7$~Oe, $H_{ac}=31$~Oe.}
\end{figure}

The dynamics of AF magnetization violates the fixed-anisotropy approximation underlying Eq.~(\ref{MT}), so the response is expected to become generally frequency-dependent. We have utilized the broadband capability of our experimental setup to analyze this dependence in the range of frequencies from a few Hz to several kHz. On the low-frequency side, this range is limited by the $1/f$ noise, and on the high-frequency side, it is limited by the properties of the ferrite core in the ac coil in our setup, whose permeability is reduced at high frequencies, especially at low temperatures. 

In measurements of the dependence on the dc field near $T_B$, ac magnetization depended on the magnetic history and evolved in time, due to the previously demonstrated magnetic aging~\cite{Ma2016}. To avoid this complication, we focus on measurements performed at a constant $H_{dc}$, after a sufficiently long delay to minimize the effects of aging. The dependence of $Re(M_T/M)$ on $H_{ac}$ was well described by Eq.~(\ref{MT}) over the entire studied ranges of temperatures and frequencies, allowing us to determine the effective field $H'$. In all these measurements, the imaginary part of ac magnetization never exceeded $15\%$ of the real part, making the distinction between the real part and the total ac magnetization insignificant. 

Figure~\ref{fig:5}(a) shows the dependence of $H'$ on frequency for four representative values of $T$, at $H_{dc}=35.7$~Oe. At $T=320$~K above $T_N$, $H'$ is nearly constant over the range $f=1-1000$~Hz. Below $T_N$, the effective field is close to its high-temperature value at small $f$, and increases with increasing frequency [see the results for $T=240$~K in Fig.~\ref{fig:4}]. This behavior is indicative of viscous AF magnetization dynamics, whose effects disappear in the low-frequency limit. At temperatures near $T_B$, the dependence of $H'$ on $f$ becomes significant, and at low frequencies it no longer approaches the high-temperature value. For instance, at $T=215$~K, $H'$ increases from $71$~Oe at $1$~Hz to $116$~Oe at $800$~Hz. We note that this variation is too large to be associated with the small unidirectional anisotropy [see Fig.~\ref{fig:3}(d)]. Instead, it likely originates from the dynamics of the canted AF magnetization, which underlies the exchange-induced uniaxial anisotropy of the ferromagnet~\cite{Schulthess98}. As the temperature is further reduced, $H'$ becomes large and independent of $f$, as illustrated in Fig.~\ref{fig:5}(a) for $T=200$~K. This behavior indicates that the AF magnetization no longer exhibits irreversible (viscous) dynamics, providing a frequency-independent exchange contribution to the effective field $H'$. 

Note that significant variations of frequency-dependent effective exchange field occur within a relatively narrow range of temperatures near $T_B$, suggesting that the irreversible AF magnetization dynamics may exhibit a critical slowdown. This hypothesis is supported by measurements of the dependence of ac magnetization on frequency at constant dc and ac fields, Figs.~\ref{fig:5}(b,c). The real part is independent of frequency at temperatures above $T_N$ and far below $T_B$. As the temperature is decreased below $T_N$ but above $T_B$, $Re(M_T/M)$ starts to decrease with increasing frequency, while still approaching its high-temperature value at small $f$ [see the $T=250$~K data in Fig.~\ref{fig:5}(b)]. The variation is most significant near $T_B$, where it is well described by the logarithmic function over the experimentally accessible frequency range. As the temperature is further reduced, the variation of $Re(M_T/M)$ with $f$ rapidly diminishes.

The central result of this paper is the observation of a qualitative change, near $T_B$, in the frequency dependence of the imaginary part of ac magnetization, providing strong evidence for the critical slowing down of AF dynamics. The value of $Im(M_T/M)$ is negligible above $T_N$ and far below $T_B$, as expected from the discussion above. As the temperature is decreased below $T_N$, an increasingly significant imaginary part emerges at high frequencies, whose magnitude monotonically decreases towards zero in the low-frequency limit, as shown for $T=250$~K in Fig.~\ref{fig:5}(c). In contrast, at $T=240$~K, $Im(M_T)$ exhibits a broad peak around $f=100$~Hz, while at $T=230$~K
the magnitude of $Im(M_T/M)$ monotonically \emph{increases} with decreasing $f$. At lower $T$, the slope of the dependence at high frequencies decreases, and the value of $Im(M_T/M)$ decreases towards zero in the low-frequency limit, as illustrated in Fig.~\ref{fig:5}(c) for $T=215$~K.

The abruptness of the variation of $Im(M_T)$ near $T_B$ is further illustrated by its dependence on temperature, Fig.~\ref{fig:5}(d). This dependence exhibits a sharp peak with the full width at half maximum of about $30$~K, with the maximum at $T=240-250$~K correlated with the onset of magnetic anisotropies associated with EB [see Fig.~\ref{fig:3}(d)]. The position of the peak does not noticeably depend on the measurement frequency, as shown for $f=100$~Hz and $1$~kHz.

\section{Discussion}

\begin{figure}
	\includegraphics[width=0.4\columnwidth]{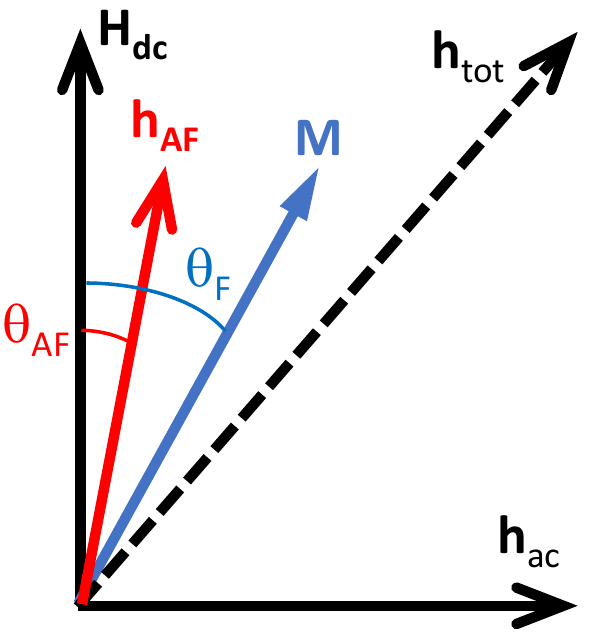} 
	\caption{\label{fig:6} (Color online). Schematic of the model for the transverse ac susceptibility incorporating the dynamical response of AF.}
\end{figure}

In this section, we discuss the implications of the observed temperature and frequency dependence of magnetic susceptibility for the magnetization state of AF, focusing mainly on the intriguing variations of the imaginary component near $T_B$, Fig.~\ref{fig:5}(c). The imaginary component of ac susceptibility is necessarily associated with irreversible dynamics leading to magnetic energy losses, as follows from the expression for the average power density delivered to the magnetic system by the ac field,
 $$
 <P>=<\frac{d\mathbf{m_T}}{dt}\cdot\mathbf{h_{ac}}>=-Im(M_T)H_{ac}.
 $$
 Such losses can be associated with irreversible domain wall motion in the inhomogeneous state of Py, as evidenced by the peaks of the imaginary component of susceptibility at the magnetization reversal points [see Figs.~\ref{fig:2}(b),~\ref{fig:3}(a)]. In contrast, the large imaginary susceptibility component that emerges close to $T_B$, and persists at high dc fields, is likely caused by the energy losses associated with irreversible processes in AF. In the simplest approximation for the polycrystalline CoO as an aggregate of weakly-coupled single-domain grains~\cite{FulcomerCharap72,OGrady2010}, such losses arise due to the irreversible transitions between the metastable states of AF magnetization. This model is equivalent to the Neel-Brown theory of magnetic viscosity due to the Arrhenius activation in single-domain magnetic nanoparticles~\cite{PhysRevB.76.054409,Egli2009}, which predicts a peak in the imaginary part of ac susceptibility at a temperature-dependent frequency 
\begin{equation}\label{NeelBrown}
f_0\approx re^{-\beta}/2\pi,
\end{equation}
where $r$ is the attempt rate, which is close to the characteristic dynamical frequency of the magnetic nanoparticle, $\beta=U/k_BT$, and $U$ is the activation energy. In more sophisticated models of F/AF bilayers that account for the elastic response of AF magnetization to exchange interaction with F, such losses are associated with the collapse of the partial AF domain walls formed due to this interaction. This leads to a similar Arrhenius activation picture [see Eq.(6) in Ref.~\cite{PhysRevB.60.12950}], predicting a peak in viscous magnetic losses at a certain temperature below $T_N$. In a completely different physical realization - the spin glasses - a peak is observed in the imaginary component of susceptibility near the glass transition temperature~\cite{RevModPhys.58.801}. Such a peak is therefore also expected to arise at the putative magnetic glass transition in the Heisenberg domain state, which was inferred from the previous time-domain measurements of magnetic aging in similar F/AF bilayers~\cite{Ma2016,PhysRevB.97.054402}.

Regardless of the specific microscopic mechanisms, to understand the implications of the complex susceptibility of F in F/AF bilayers, it is necessary to establish the relationship between this quantity and the dynamical response of AF. The main complication is that the susceptibility of F only indirectly reflects the properties of AF. In particular, it remains finite even if the effects of AF are negligible. In Section~\ref{experiment}, we analyzed  the transverse ac susceptibility in the static-anisotropy approximation, corresponding to the stable magnetization state of AF. Our analysis predicted that the ac magnetization oscillates in-phase with the ac field, with the amplitude independent of frequency. Therefore, to account for the observed non-negligible imaginary component of susceptibility and its dependence on frequency, the non-static nature of the magnetization state of AF must be incorporated in the analysis.

We focus on the small-amplitude oscillations of quasi-uniform states, formed by the magnetization of Py at sufficiently high dc fields. In this limit, all the contributions to the magnetization dynamics can be described in terms of effective time-dependent fields representing the corresponding contributions to the instantaneous gradient of the magnetic energy of Py with respect to the orientation of it magnetization [Fig.~\ref{fig:6}]. For the driving ac field described by $h_{ac}(t)=H_{ac}cos\omega t$, the small-angle oscillation of the Py magnetization $\mathbf{M}$ is described by the time-dependent angle $\theta_F(t)=Re\theta_F\cos(\omega t)-Im\theta_F\sin(\omega t)$. The effect of exchange interaction at the F/AF interface can be described in terms of the time-dependent effective exchange field $\mathbf{h_{AF}}$, forming an angle $\theta_{AF}(t)=Re\theta_{AF}\cos(\omega t)-Im\theta_{AF}\sin(\omega t)$ with respect to the dc field. In the small-angle limit $H_{ac}\ll H_{dc}$, $\theta_F,\theta_{AF}\ll 1$. 

To analyze the dynamics of AF magnetization, described by the time-dependent vector $\mathbf{h_{AF}}$, we note that losses associated with the Arrhenius activation in the Neel-Brown-type models~\cite{FulcomerCharap72,PhysRevB.60.12950,OGrady2010} can be described in terms of the magnetic viscosity~\cite{Ediger1996}. One can expect that the magnetic viscosity should also describe the dynamical response of the putative magnetic glass in the Heisenberg domain state. Regardless of the microscopic nature of the AF state, we can describe its periodic dynamics around $\theta_{AF}=0$, driven by the exchange interaction with F, in terms of the sum of an elastic and a viscous contribution,
\begin{equation}\label{dAF}
K\theta_{AF}+\nu\frac{d\theta_{AF}}{dt}=\theta_F-\theta_{AF},
\end{equation}
where $K$ is the elastic coefficient defined by the static magnetic properties of AF and the effective exchange field at the F/AF interface, and $\nu$ is the effective viscosity in units of time, determined by the energy relaxation processes in AF. In the Neel-Brown theory, $\nu$ is the thermal activation time~\cite{Egli2009}. The magnetization state of the ferromagnet is determined by the balance of the torques exerted on $\mathbf{M}$,
\begin{equation}\label{dF}
H_{AF}(\theta_F-\theta_{AF})+H_{dc}\theta_F=H_{ac}cos(\omega t).
\end{equation}

In our measurements, the imaginary part of the transverse magnetization was always significantly smaller than its real part. Neglecting its effect on the AF dynamics in Eq.~(\ref{dAF}), we obtain
\begin{equation}\label{thetaAF}
\begin{split}
Re\theta_{AF}&=(K+1)Re\theta_F/L(\omega),\\
Im\theta_{AF}&=\nu\omega Re\theta_F/L(\omega),
\end{split}
\end{equation}
where $L(\omega)=(K+1)^2+(\nu\omega)^2$. Plugging this result into Eq.~(\ref{dF}), we obtain
\begin{equation}\label{thetaF}
\begin{split}
Re\theta_F&=\frac{H_{ac}}{H_{dc}+H_{AF}[1-(K+1)/L(\omega)]}\\
Im\theta_F&=-\frac{\nu\omega H_{AF}Re\theta_F}{(H_{dc}+H_{AF})L(\omega)},
\end{split}
\end{equation}
describing the ac magnetization $M_T/M\approx \theta_F$ measured in our experiment at small $H_{ac}$.

We start our analysis of Eqs.~(\ref{thetaF}) by considering simple limiting cases. At high temperatures $T>T_N$, the effects of AF should be negligible, corresponding to $H_{AF}=0$. In this limit, $Re\theta_F=H_{ac}/H_{dc}$ and $Im\theta_F=0$, as expected for a standalone F. In another limit discussed in Section~\ref{experiment}, according  to Eq.~(\ref{thetaAF}), the dynamics of AF becomes negligible at $\nu\omega\gg 1$, since $L(\omega)\to\infty$ in this limit. Equation~(\ref{thetaF}) then gives $Re\theta_F=H_{ac}/(H_{dc}+H_{AF})$, $Im\theta_F=0$, consistent with Eq.~(\ref{MTlin}). In the zero-frequency limit, the viscous contribution to Eq.~(\ref{thetaF}) vanishes, giving $Re\theta_F=\frac{H_{ac}}{H_{dc}+H_{AF}(1-1/(K+1))}$, $Im\theta_F=0$. This response is only weakly dependent on the value of $K\ge 0$, so in the calculations discussed below we use $K=0$.

\begin{figure}
	\includegraphics[width=1\columnwidth]{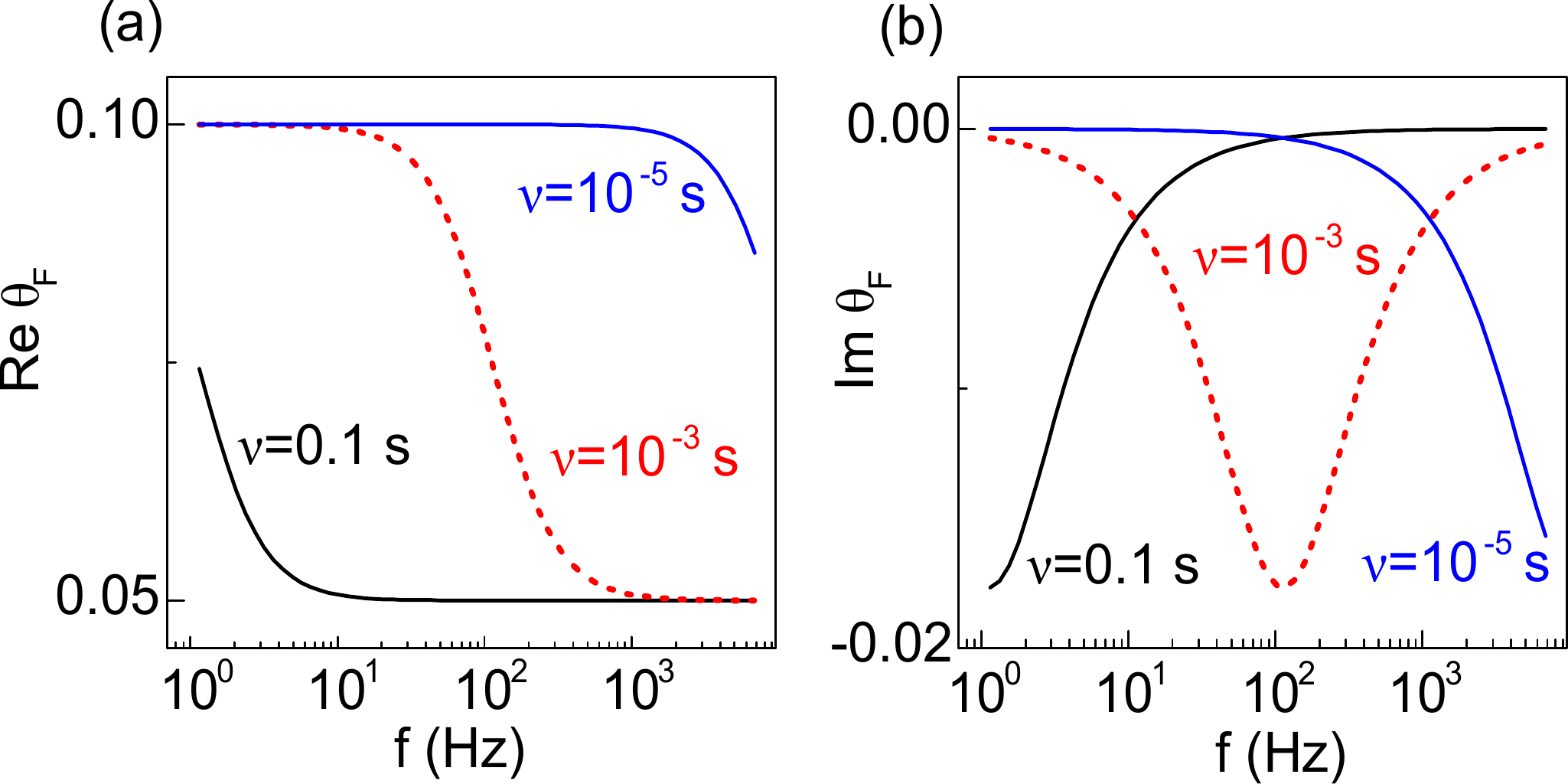} 
	\caption{\label{fig:7} (Color online). Real (a) and imaginary (b) parts of $\theta_F$ vs $f$, calculated using Eq.~(\ref{thetaF}), at the labeled values of viscosity $\nu$. The calculations were performed with $H_{dc}=H_{AF}=10 H_{ac}$, $K=0$.}
	\end{figure}

Figure~\ref{fig:7} shows the calculated $\theta_F(f)$ for the range of viscosities that result in a significant imaginary susceptibility component at the experimental frequencies $f=1$~Hz-$7$~kHz. The real part of $\theta_F$ monotonically increases with decreasing frequency, with the largest variation at $\nu\approx1/2\pi f_0$, where $f_0\approx 10^2$~Hz is the characteristic measurement frequency. Comparing with the data in Fig.~\ref{fig:5}(b), we conclude that the viscosity increases with decreasing temperature, with $\nu\approx 10^{-2}$~s at $T=240$~K. Because of the inhomogeneities in the studied system associated with the variations of the local thickness, magnetic anisotropy, and the level of frustration introduced by the exchange interactions at F/AF interface, we do not expect that a single value of $\nu$ is sufficient to quantitatively describe the behaviors at a given temperature. This is reflected by the significantly more gradual experimentally observed variation of $Im(M_T/M)$ with frequency.

According to Eqs.~(\ref{thetaF}), the imaginary part of response should vanish in both the low- and the high-frequency limits, with a peak at $\nu\omega=(K+1)$. 
At $\nu=0.1$~s and $K=0$, this peak is below the frequency scale in Fig.~\ref{fig:7}(b), resulting in a monotonic decrease of $Im\theta_F$ with increasing $f$. In contrast, at  $\nu= 10^{-3}$~s, the peak shifts close to $f=100$~Hz in the middle of the frequency range, while at $\nu=10^{-5}$~s, the peak shifts above $7$~kHz, resulting in the monotonic increase of $Im(\theta_F)$ with increasing $f$. These results are in a good qualitative agreement with the data in Fig.~\ref{fig:5}(c) for $T=230$~K, $240$~K, and $250$~K, respectively, suggesting that the effective viscosity of AF increases by four orders of magnitude over this temperature range. For Arrhenius activation, such a rapid variation of viscosity places very stringent constraints on the distribution of the activation barriers and the possible values of the activation attempt rates. For the studied thin-film Py/CoO bilayers, previously shown to exhibit non-Arrhenius cooperative magnetization dynamics~\cite{Ma2016}, our observation of such a rapid viscosity variation is consistent with the magnetic freezing expected to occur at the putative magnetic glass transition in the Heisenberg domain state~\cite{PhysRevB.97.054402}.

\section{Conclusions}

The main technical result presented in this paper is the demonstration that the variable-temperature, variable-frequency transverse ac susceptibility technique can provide detailed information about the anisotropy and dynamical properties of magnetic systems. In particular, we showed that the uniaxial and unidirectional anisotropies can be determined from the dependence of susceptibility on the dc field, and independently from the nonlinear dependence on the ac field. Furthermore, the dependence of the real and imaginary parts of susceptibility on frequency and temperature provide information about the magnetic energy relaxation, and ultimately about the nature of the magnetization state of the system.

We utilized the transverse ac susceptibility technique to analyze the magnetic characteristics of a common exchange bias system - a thin-film CoO/Permalloy bilayer. We found that the exchange bias - asymmetry of the hysteresis loop with respect to the field direction - arises in the studied system mainly due to the different reversal mechanisms between the two opposite magnetization directions stabilized by the large uniaxial anisotropy. The latter is likely associated with the reversible canting (partial spin-flop) of the antiferromagnetic moments due to the exchange interaction with the ferromagnet. These findings may provide a significant step towards general fundamental understanding of phenomena associated with exchange interactions at the ferromagnet/antiferromagnet interfaces, and enable engineering of these phenomena for applications in nanomagnetic devices. A number of significant, experimentally verifiable consequences of these observations are expected. For instance, the magnetization reversal in sufficiently small nanomagnets proceeds through quasi-uniform magnetization rotation. Therefore, the asymmetry of the hysteresis loop in nanoscale F/CoO bilayers is expected to become significantly reduced, while the coercivity is expected to become enhanced by up to a factor of two. Furthermore, since the magnetization reversal into the exchange bias direction is likely controlled by the inhomogeneous exchange torques at F/AF interface, exchange bias is not expected to scale inversely  with the thickness of the ferromagnet, because the effects of inhomogeneity become averaged out over the lengthscales defined by the ferromagnet's thickness. Finally, the anisotropy of the ferromagnet itself is expected to provide a non-trivial contribution to exchange bias, because it affects the domain structure and the characteristics energy scale of domain walls.

Our third main result is the observation of a rapid variation of magnetic viscosity with temperature close to the exchange bias blocking, marked by a sharp peak in the temperature dependence of the imaginary part of transverse ac susceptibility. This result is consistent with the conjecture, previously put forward based on the aging measurements, that the Heisenbeg domain state in thin-film  ferromagnet/antiferromagnet bilayers is a magnetic glass~\cite{PhysRevB.97.054402}.

The ability to judiciously generate magnetic liquid and magnetic glass states in thin-film heterostructures can become useful both for the fundamental research and for applications. The glass state of matter has been one of the most extensively researched subjects in condensed matter physics over the last several decades, but a universally accepted theory of this state has not yet emerged~\cite{Ediger1996}. Experimental studies of physical glasses have been hindered by the difficulty in precisely replicating the studied systems, and in identifying the physical characteristics most relevant to the glass formation. Magnetic and spin glasses are straightforward to characterize and model, and the spin system can be "reset" by heating without modifying any other parameters, providing the ability of repeated measurements on identical systems~\cite{RevModPhys.58.801}. However, experimental studies have been largely limited to bulk dilute spin systems at low temperatures, making it difficult to address important questions related to the role of dimensionality of the material and of the order parameter~\cite{Proctor2014}, the effects of surfaces/interfaces, and interactions competing with the glass transition. 

Spin systems exhibiting controllable complex collective behaviors at room temperature can provide a suitable medium for a number of applications actively discussed by the soft condensed matter community~\cite{PhysRevLett.112.198001,PhysRevLett.120.224101}. For instance, the stability of a multitude of configurations, determined both by the strength of the perturbations and by their history, may lend itself to efficient neuromorphic circuit implementations. We note that in contrast to ferromagnets, the density of information in magnetic glasses is not limited by the competition among the local exchange, dipolar, and thermal energies. Furthermore, while glassy systems are generally slow, they can also posses fast degrees of freedom that can extend beyond typical dynamical frequencies of ferromagnets, due to the relaxed angular momentum conservation conditions. Thus, such systems can share some of the advantages of AF spintronics, while providing more straightforward possibilities for the control and detection of their states.

\begin{acknowledgements}
We acknowledge support from the NSF Grant DMR-1504449, and thank Connie Roth and Justin Burton for the helpful discussions.
\end{acknowledgements}
\bibliography{CoOac}{}
\bibliographystyle{apsrev4-1}

\end{document}